\documentclass[aps,prl,twocolumn,amsmath,amssymb,superscriptaddress,floatfix,longbibliography]{revtex4-2}
\pdfoutput=1 
\usepackage{tabularx}
    \newcolumntype{L}{>{\raggedright\arraybackslash}X}
\usepackage{makecell}
\usepackage{caption}
\usepackage{xcolor}
\usepackage{multirow}
\usepackage{units}
\usepackage{eucal}
\usepackage{bbm}
\usepackage{physics}
\usepackage{import}
\usepackage{graphicx}
\usepackage{tikz}
\usepackage{rotating}
\usepackage{subcaption}
\usepackage[unicode=true,pdfusetitle,
bookmarks=true,bookmarksnumbered=false,bookmarksopen=false,
 breaklinks=false,pdfborder={0 0 1},backref=false,colorlinks=true,citecolor=blue,
 urlcolor=blue,linkcolor=blue]{hyperref}
\usepackage[mode=buildnew]{standalone}
\usepackage{lmodern}

\newcommand{\channel}{\mathcal{M}}
\newcommand{\shadowtable}{T}

\newcommand{\volume}{V}
\newcommand{\pairings}[1]{P_{#1}}
\newcommand{\pairing}{\pi}
\newcommand{\body}{{w}}
\newcommand{\nz}{{n_z}}
\newcommand{\np}{{n_+}}

\renewcommand{\op}{O}
\newcommand{\opstring}{S}
\newcommand{\canopstring}{\opstring_{\circ}}

\newcommand{\ampeigmat}{G}
\newcommand{\ampspread}[1]{\alpha_{#1}}
\newcommand{\invampspread}[1]{\beta_{#1}}

\newcommand{\Z}{Z}
\newcommand{\plus}{a^{\dagger}}
\newcommand{\minus}{a}

\newcommand{\shadownorm}[1]{\|{#1}\|_{\text{s}}}

\renewcommand{\vec}[1]{\boldsymbol{#1}}

\begin{document}

\title{Efficient Local Classical Shadow Tomography with Number Conservation}

\author{Sumner N. Hearth}
\author{Michael O. Flynn}
\author{Anushya Chandran}
\author{Chris R. Laumann}
\affiliation{%
Department of Physics, Boston University, 590 Commonwealth Avenue, Boston, Massachusetts 02215, USA
}%

\newcommand{\refa}[1]{{\color{red} Ref1: #1}}
\newcommand{\refb}[1]{{\color{blue} Ref2: #1}}

\date{\today}

\begin{abstract}
Shadow tomography aims to build a classical description of a quantum state from a sequence of simple random measurements.
Physical observables are then reconstructed from the resulting classical shadow.
Shadow protocols which use single-body random measurements are simple to implement and capture few-body observables efficiently, but do not apply to systems with fundamental number conservation laws, such as ultracold atoms. 
We address this shortcoming by proposing and analyzing  a new local shadow protocol adapted to such systems. 
The ``All-Pairs'' protocol requires one layer of two-body gates and only $\text{poly}(\volume)$ samples to reconstruct arbitrary few body observables.
Moreover, by exploiting the permutation symmetry of the protocol, we derive a linear time post-processing algorithm which applies to both hardcore bosons and spinless fermions in any spatial dimension.
We provide a proof-of-principle reference implementation and demonstrate the reconstruction of 2- and 4-point functions in a paired Luttinger liquid of hardcore bosons.
\end{abstract}

\maketitle

\begin{figure}
    \centering
    \includegraphics[width=0.9\columnwidth]{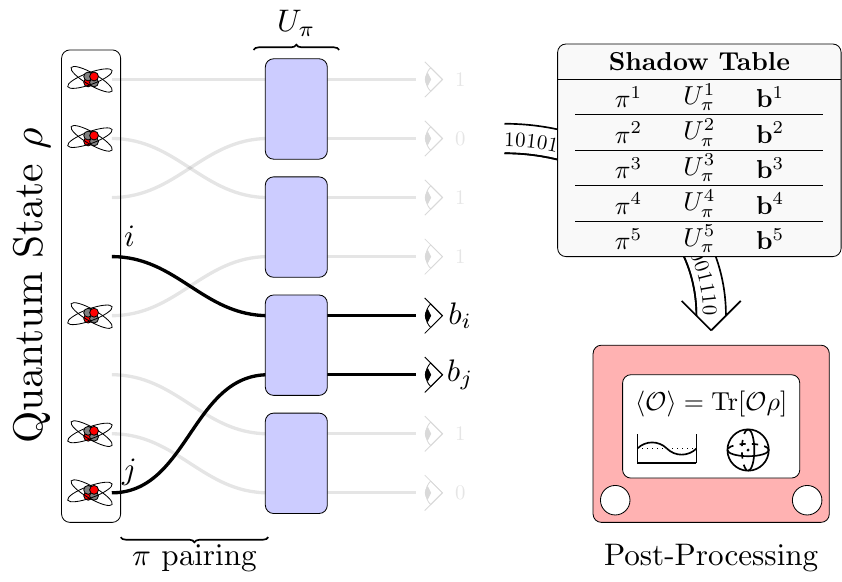}
    \caption{The All-Pairs shadow protocol: 
    for each copy of the quantum state $\rho$, (1) randomly choose a pairing $\pairing$ of the $\volume$ sites and apply swaps to bring paired sites together. The figure shows a swap circuit implementing the pairing $\pairing = [1 3][2 5][4 8][6 7]$.  (2) Apply random 2-body number-conserving unitary gates to each pair $[ij]$ in $\pairing$. (3) Measure  in the occupation basis with result $\mathbf{b}$. 
    (4) Record the chosen pairing, gates, and outcomes as a row in the shadow table.
    Few-body observables can be efficiently re-constructed from the shadow table.}
    \label{fig:allpairscartoon}
\end{figure}

Quantum state tomography aims to produce a complete classical description of the state $\rho$ of a quantum system: a prohibitive task requiring exponentially many measurements of independently prepared copies of $\rho$. 
Rather than measure all possible matrix elements, recent works have taken a statistical approach designed to capture classes of observables of physical interest~\cite{AaronsonShadows,AaronsonGentleMeasurements, ManyPropertiesFewMeasurements,Shivam_2023,Liu_Hao_Hu_2023,Chen_Yu_Zeng_Flammia_2021,Zhao_2021,Koh_2022,EntanglementHamiltonianZoller,ShadowScrambling,LearningConservationLaws,LocallyBiasedShadows,ProcessTomography,BarrenPlateausAndShadows,Zhou_Zhang_2023,Garratt_Altman_2023}.
The shadow tomography framework is illustrated in Fig.~\ref{fig:allpairscartoon}: for each copy of $\rho$, sample a unitary circuit $U$ from a bespoke ensemble, apply it to $\rho$, and measure in the computational basis. 
Physical observables can later be reconstructed by classical post-processing from the record of applied unitaries and measurement outcomes -- the \emph{classical shadow}.

\begin{figure}[t]
    \centering
    \begin{subfigure}[t]{0.6\columnwidth}
        \centering
        \vfill
        \includegraphics[width=0.8\linewidth]{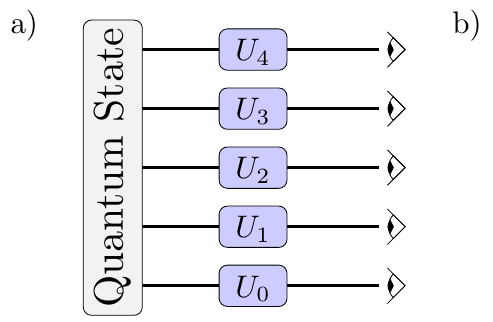}
    \end{subfigure}\hspace{-0.25cm}
    \begin{subfigure}[t]{0.4\columnwidth}
        \centering
        \vfill
        \bgroup
        \def\arraystretch{1.2}
        \begin{tabular}{c|c}
            $\opstring_i \opstring_j$ & $\channel_{[ij]}[\opstring_i \opstring_j]$\\ \hline\hline
            $I_i I_j$ & $I_iI_j$ \\
            $\Z_i\Z_j$ & $\Z_i\Z_j$ \\\hline
            $I_i\Z_j$ & $\frac{2}{3}I_i\Z_j + \frac{1}{3}\Z_iI_j$ \\
            $\Z_iI_j$ & $\frac{2}{3}\Z_iI_j + \frac{1}{3}I_i\Z_j$ \\\hline
            $\plus_i \minus_j$ & $\frac{1}{3}\plus_i \minus_j$ \\
            $\minus_i \plus_j$ & $\frac{1}{3}\minus_i \plus_j$ \\\hline
            \text{Else} & 0
        \end{tabular}
        \egroup
    \end{subfigure}
    \caption{a) A product protocol of single-body unitaries.  b) The action of $\channel_{[ij]}$ defined in Eq.~\eqref{Eq:Mijdefinition} on the operator basis. The channel preserves the number of $I$, $\Z$, $\plus$, and $\minus$ operators, but can rescale/swap operators.}
    \label{fig:channelaction}
\end{figure}

Perhaps the most important shadow protocol is adapted to the efficient reconstruction of few-body observables.
The \emph{product} protocol is deceptively simple: each $U$ is a product of independently sampled random 1-body gates (see Fig.~\ref{fig:channelaction}a).
The protocol is efficient by several measures (cf. row 1 of Table~\ref{tab:channelsummary}):
each measurement requires only a few quantum gates (low \emph{gate complexity}), while the number of classical post-processing steps scales at most linearly with the number of qubits $\volume$ (low \emph{classical complexity}). 
The number of samples required to estimate a few-body observable scales exponentially with the support $\body$ of the observable, but is  independent of system size (low \emph{sample complexity}). 
Accordingly, the product protocol has developed into an important experimental tool for characterizing many-qubit systems~\cite{BeenakkerRandom,BrydgesRandomMeasurements,ElbenStatisticalToolbox,NatureShadowReview,Ippoliti_2023}.

Many quantum systems, such as ultracold atoms~\cite{UltracoldReview1,UltracoldReview2,Kaufman_Lester_Reynolds_Wall_Foss-Feig_Hazzard_Rey_Regal_2014}, are constrained by fundamental symmetries which restrict the available unitary gates as well as the measurement basis. 
Such restrictions render the simple product protocol tomographically incomplete.
As an example, consider an atom living in two sites. 
The only single-site gates which are consistent with the conservation of atom number are phase operators.
These do not affect the statistics of measurements in the occupation basis.
Consequently, no product shadow protocol can distinguish the two Bell states $\ket{01} \pm \ket{10}$.

Of course, \emph{two}-body gates are sufficient to rotate the Bell basis into the occupation basis and number-conserving shadow protocols based on global random unitaries and/or deep quantum circuits are generically tomographically complete~\cite{ManyPropertiesFewMeasurements,Tran_Mark_Ho_Choi_2023,Bringewatt_Kunjummen_Mueller_2023}.
However, as such unitaries scramble local information, these protocols lose any advantage for reconstructing few-body observables: the sample complexity scales exponentially in system size.

In this article, we propose a tomographically complete shadow protocol adapted to reconstructing few-body observables in number-conserving systems of fermionic or bosonic hard-core atoms.
The ``All-Pairs'' protocol is straightforward  (see Fig.~\ref{fig:allpairscartoon}):
the random unitaries are constructed by first choosing a random pairing between all sites in the system and then choosing an independent random two-body gate on each pair.
It turns out that this shallow but fluctuating circuit geometry permits access to arbitrary few-body observables without scrambling local information.
More precisely, the sample complexity scales polynomially with system size $\volume$.

Unlike global channels and deep circuits, the All-Pairs protocol has low classical complexity (scaling linearly with $\volume$), so that the reconstruction of observables is efficient. 
This efficiency is a consequence of the permutation symmetry of the protocol, which leads to significant analytic reductions.
Table~\ref{tab:channelsummary} presents analytically obtained complexity bounds for the All-Pairs protocol, and compares it to other protocols discussed in the literature~\cite{ManyPropertiesFewMeasurements,Hu_Choi_You_2023,Low_2022,Tran_Mark_Ho_Choi_2023,Mark_Choi_Shaw_Endres_Choi_2023}.

We first review key concepts from the shadow tomography formalism.
Using this formalism, we express the All-Pairs protocol as a quantum channel and analyze its decomposition into symmetry sectors.
The eigenvalues and eigenoperators of the channel bound the classical and sample complexity for the protocol.
Finally, as a proof-of-concept, we reconstruct 2- and 4-point functions in a bosonic model with a paired Luttinger liquid ground state.

\paragraph{Background and Notation ---}
We adopt the shadow tomography formalism developed in Ref.~\cite{ManyPropertiesFewMeasurements}. 
The \emph{shadow channel} describes the action of the shadow measurement protocol, 
\begin{equation}
\begin{aligned}
    \channel[\rho] &= \int dU \sum_{\vec{b}} \Tr[\rho \left( U^\dagger \ketbra{\vec{b}}{\vec{b}} U\right) ] \left( U^\dagger \ketbra{\vec{b}}{\vec{b}} U\right)\\
    &\equiv \mathop{\mathbb{E}}\limits_{U, \vec{b}}\left[ U^\dagger \ketbra{\vec{b}}{\vec{b}} U \right],
\end{aligned}
\end{equation}
on a quantum state $\rho$. 
Here, $U$ is the randomly sampled unitary and $\ket{\vec{b}}$ is the post-measurement basis state.
The shadow protocol is tomographically complete if and only if the channel is invertible, in which case 
\begin{align}
    \rho
    &= \mathop{\mathbb{E}}\limits_{U, \vec{b}}[\channel^{-1}[U^\dagger \ketbra{\vec{b}}{\vec{b}} U]] . \label{eq:ChannelExample}
\end{align}
The shadow table $\shadowtable$ is the set of independently sampled $U$ and obtained measurement outcomes $\vec{b}$.
An unbiased estimator $\hat{\rho}$ for $\rho$ is constructed by replacing the $\mathbb{E}$ in Eq.~\eqref{eq:ChannelExample} with an empirical average over the entries in $\shadowtable$. 
The application of $\channel^{-1}$ is done classically, so that the difficulty of the inversion determines the post-processing complexity.

Given the inverse channel, estimating the expectation of an operator $\langle \op \rangle$ from a measurement record is straightforward,
\begin{align}
    \widehat{\langle \op \rangle} = 
    \Tr[\hat\rho \op ] &= \frac{1}{|\shadowtable|} \! \sum_{(U,\vec{b}) \in \shadowtable} \!\!\! \Tr[\left(U^\dagger \ketbra{\vec{b}}{\vec{b}} U\right)\channel^{-1}[\op]] 
    \label{eq:estimatordescription}
\end{align}
Here, we have used the hermiticity of $\channel^{-1}$ with respect to the trace inner product. 
Clearly, the number of samples in $\shadowtable$ and the variance of the underlying distribution determine the statistical error on the estimate of the expectation value.
The number of samples required to obtain a fixed standard error is controlled by the \emph{shadow norm} of the operator $\op$~\footnote{We include an absolute value compared to \cite{ManyPropertiesFewMeasurements} to account for non-Hermitian operators $\op$. This amounts to controlling the real and imaginary parts of the estimator.},
\begin{align}
    \shadownorm{\op}^2 = \max\limits_{\rho: \text{ state}} ~ \mathop{\mathbb{E}}\limits_{U, \vec{b}} \left| \Tr[\left(U^\dagger \ketbra{\vec{b}}{\vec{b}} U\right) \channel^{-1}[\op]] \right|^2 . \label{eq:shadownorm}
\end{align}
The RHS is maximized over density matrices $\rho$, and as noted in \cite{ManyPropertiesFewMeasurements}, only the traceless components of $O$ contribute to the shadow norm.
In practice, one employs \emph{median-of-means} estimation to reduce the probability of large errors~\cite{ManyPropertiesFewMeasurements}.

Although the protocol applies to fermions as well, we focus on a system of conserved hard-core bosons on $\volume$ sites and adopt a local operator basis adapted to number conservation, $\opstring \in \{I, \Z, \plus, \minus\}^{\otimes \volume}$ where $\plus=\frac{1}{2}(X-iY)$ and $\minus=\frac{1}{2}(X+iY)$ are the usual raising and lowering operators for a 2-level system, with the $\ket{\uparrow}$ state identified as the empty state.
Number-conserving strings are those in which $\plus$ and $\minus$ appear in equal number.
The number of $\plus$ or $\minus$ operators in a basis string is denoted by $\np$, and the number of $\Z$ operators by $\nz$.
It is straightforward to confirm these strings form an orthogonal basis for all number-conserving operators on a $\volume$-site system (Supp.~Ia~\cite{supplemental}).

\bgroup
\def\arraystretch{1.15}
\begin{table*}[!ht]
    \centering
    \begin{tabular}{c|c|c|c|c|c}
        \multirow{2}{*}{Protocol} & \multirow{2}{*}{\makecell{U(1)\\ Compatible}} & \multirow{2}{*}{\makecell{Gate\\ Complexity}} & \multicolumn{2}{c|}{Classical Complexity} & \multirow{2}{*}{\makecell{Sample\\ Complexity}} \\
         & & & \makecell{One-Time} & \makecell{Per Sample} &  \\ \hline \hline
        Product\cite{ManyPropertiesFewMeasurements}  & No & $\volume$ & & $\body$ & $3^w$ \\ \hline
        Global Random Circuit\cite{ManyPropertiesFewMeasurements} & Yes & $\volume^{2}$ \cite{Hearth_Flynn_Chandran_Laumann_2023} &  & $\volume^{2} 2^{\volume}$ & $2^{\volume}$ \\
        Ergodic Evolution\cite{Tran_Mark_Ho_Choi_2023} & Yes & Hamiltonian & $\left(2^{2\volume}\right)^3$ & $2^{2\volume}$ & $\ge 2^\volume$ \\
         \hline
        All-Pairs & Yes & $\frac{\volume}{2}$ & $(\nz)^3$ & $\volume + \nz 2^{2\nz}$ & $\left(\frac{3}{2} \right)^{\np+2\nz} \times \frac{\volume^{\np}}{\np!}$ \\
    \end{tabular}
    \caption{Resource complexity for estimating the expectation value of $\body$-body operator strings allowing long range few-body operators. The ergodic protocol has a sample complexity which depends on the native Hamiltonian and evolution time.}
    \label{tab:channelsummary}
    \end{table*}
\egroup

\paragraph{The All-Pairs channel ---}

We now apply the shadow formalism to the All-Pairs protocol.
The channel is an average over pairings $\pairing$ from the set of all possible pairings of sites $\pairings{\volume}$,
\begin{align}
    \channel &= \frac{1}{|\pairings{\volume}|} \sum_{\pairing \in \pairings{\volume}} \channel_{\pairing} . \label{eq:allpairdefm}
\end{align}
Each pairing $\pairing$ is a grouping of the $\volume$ sites into $\volume/2$ pairs.
The number of such pairings is $|\pairings{\volume}| = (\volume - 1)!!$.
For odd $\volume$ a single site is left out of any given pairing and $|\pairings{\volume}| = \volume (\volume - 2)!!$.
Each pair of sites $[ij] \in \pairing$ is acted on by 2-body unitaries, resulting in a product channel
\begin{equation}
\begin{aligned}
    \channel_{\pairing} &= \prod_{[ij] \in \pairing} \channel_{[ij]} , \\
    \channel_{[ij]}[\rho] &= \mathop{\mathbb{E}}\limits_{U_{[ij]}, b_i b_j}\left[U^\dagger_{[ij]} \ketbra{b_i b_j}{b_i b_j} U_{[ij]}\right] .\label{Eq:Mijdefinition}
\end{aligned}
\end{equation}
Sampling $U_{[ij]}$ from an ensemble which forms a 2-design over the number-conserving unitary group~\cite{Marvian_2022,Hearth_Flynn_Chandran_Laumann_2023,Weingarten1978,Collins:2022aa}, $\channel_{[ij]}$ can be evaluated block-by-block,
\begin{align}\label{eq:SinglePairAction}
    \channel_{[ij]}\left[\begin{pmatrix}
        \rho_0 &  \\
        & \rho_1 & \\
        & & \rho_2
    \end{pmatrix}\right] &= \begin{pmatrix}
        \rho_0 &  \\
        & \frac{1}{3}\left(\rho_1 + I_{2\times 2} \Tr \rho_1\right) & \\
        & & \rho_2
    \end{pmatrix} .
\end{align}
Here, $\rho_0$ and $\rho_2$ are scalars representing the $m=0$ and $m=2$ number sectors, whereas $\rho_1$ is a $2\times 2$ matrix describing states $\ket{01}$ and $\ket{10}$ in the occupation basis.
Expanding Eq.~\eqref{eq:SinglePairAction} in the basis of operator strings yields the table in Fig.~\ref{fig:channelaction}b.

\paragraph{Analysis of the Channel---}
It is straightforward to show that the All-Pairs channel is tomographically complete.
Consider an operator string $\opstring$ we wish to reconstruct using the information in the shadow table $\shadowtable$.
Begin by filtering $\shadowtable$ down to only those rows whose pairings match $\plus$ with $\minus$ and $\Z$ with another $\Z$ operator within $\opstring$; call this filtered channel $\channel_\opstring$.
It is clear from Fig~\ref{fig:channelaction}b that the action of this effective channel is diagonal on $\opstring$ and inversion is trivial $\channel^{-1}_\opstring [\opstring] = 3^{\np}\opstring$.
The choice of $\opstring$ is arbitrary so all operators can be reconstructed from the information contained in $\shadowtable$. 
However, the process described above is inefficient and throws out most entries of the table.
To make the protocol more efficient and greatly reduce its sample complexity we now diagonalize the channel without prior filtering.

The All-Pairs channel can be efficiently diagonalized and inverted due to strong symmetry constraints.
First, $\channel$ is manifestly permutation symmetric. 
For any permutation $\sigma$ of the $\volume$ sites,
\begin{align}
    \channel[\sigma \opstring \sigma^{\dagger}] &= \sigma \channel[ \opstring] \sigma^{\dagger} .
\end{align}
Second, $\channel$ conserves the number of $I,\Z,\minus$ and $\plus$  operators in an operator string $\opstring$.
Third, $\plus$ and $\minus$ operators are immobile under the action of $\channel$ (cf. Fig.~\ref{fig:channelaction}b).

Consider an operator string with $\np$ instances of $\minus$ and $\plus$ and $\nz$ instances of $\Z$ in a canonically ordered form, $\canopstring~=~\left(\plus\otimes\minus\right)^{\otimes\np}\otimes \Z^{\otimes\nz} \otimes I^{\otimes \volume - \body}$.
Since $\channel_\pairing[\canopstring]$ vanishes unless $\pairing$ pairs the $\minus$ and $\plus$ operators, the channel action factorizes, 
\begin{equation}
\begin{aligned}
\channel[\canopstring] 
&= \frac{f}{3^{\np}} (\plus\otimes\minus)^{\otimes \np} \otimes \channel\left[\Z^{\otimes \nz} \otimes I^{\otimes \volume - \body}\right] . \label{eq:channelsep}
\end{aligned}
\end{equation}
Here, $f$ denotes the fraction of pairings which contribute,
\begin{align}
    f &= \np! \frac{|\pairings{(\volume - 2 \np)}|}{|\pairings{\volume}|} . \label{eq:pairingfraction}
\end{align}

We now focus on the action of $\channel$ restricted to the space of operators spanned by $S \in \{I,\Z\}^{\otimes \volume}$. 
Since $\channel$ conserves the number of $Z$'s, it defines a permutation-symmetric hard-core random walk of the $Z$'s (formally, a symmetric exclusion process). 
Accordingly, the channel decomposes into $Z$-sectors, labeled by $\nz$, and irreducible representations of the permutation group, labeled by $\lambda = (\volume - \lambda_2, \lambda_2)$.
Each $\nz, \lambda$ irrep appears once and has a corresponding eigenvalue $c_\lambda$ which is independent of $\nz$.

We use combinatorial arguments to derive closed but rather long expressions for the eigenvalues $c_\lambda$ in Supp.~IIc~\cite{supplemental}. 
This decomposition underlies the computationally efficient inversion scheme and the rigorous bounds on the sample complexity.

\paragraph{Inverting the Channel---}

To compute the estimator Eq.~\eqref{eq:estimatordescription} from the shadow table, we need an efficient algorithm to apply the inverse channel $\channel^{-1}$ to string operators within the trace.
The description of the reconstruction algorithm is presented in Supp.~II~\cite{supplemental}. 
Here, we sketch the key steps which reduce the problem to inverting an $(\nz + 1) \times (\nz + 1)$ matrix. 

With reference to Eq.~\eqref{eq:channelsep}, the non-trivial part of inverting $\channel$ comes from its action on strings $\opstring \in \{I,\Z\}^{\otimes \volume}$.
Permutation symmetry and $n_z$ conservation impose that
\begin{align}
    \channel[\canopstring] &= \sum_{d=0}^{\nz} \ampspread{d} \sum_{\substack{\opstring_d \text{ at}\\ \text{distance }d}} \opstring_d , \label{eq:ampspreaddist}
\end{align}
where the strings $S_d$ are at swap distance $d$ from the reference string $\canopstring$. 
Using \eqref{eq:allpairdefm}, it is possible to obtain the amplitudes $\alpha_d$ in closed form (Supp.~IIb~\cite{supplemental}). 

The two decompositions, into $c_\lambda$ and into $\alpha_d$, linearly parametrize the channel $\channel$ in a given $n_z$ sector. Thus, there is a linear relationship,
\begin{align}
    c_\lambda = \sum_{d=0}^{\nz} \ampeigmat_{\lambda d}\ampspread{d}
    \label{eq:linearrelation}
\end{align}
where $\ampeigmat$ is an $(\nz+1) \times (\nz+1)$ integer matrix determined entirely by the symmetry of the channel.
We use combinatorial arguments to obtain $G$ in closed form in Supp.~IIc~\cite{supplemental}.

The inverse channel is simple in terms of its irrep decomposition, its eigenvalues are simply $\frac{1}{c_\lambda}$. 
In order to efficiently apply $\channel^{-1}$ to a string $S$, we need the amplitudes $\beta_d$ which govern how it delocalizes $S$ to other strings with swap distance $d$. 
These can be obtained using the inverse of Eq.~\eqref{eq:linearrelation}. 
Schematically, we perform the following steps,
\begin{align*}
    \channel \rightarrow \ampspread{d} \xrightarrow{~\ampeigmat~} c_{\lambda} \xrightarrow{\text{invert}} \frac{1}{c_{\lambda}} \xrightarrow{\ampeigmat^{-1}} \invampspread{d} \rightarrow \channel^{-1}
\end{align*}
Eq.~\eqref{eq:estimatordescription} thus becomes,
\begin{align}
    \!\!\!\!\!\Tr[\hat{\rho} \canopstring] &= \frac{1}{|T|} \!\! \sum_{(U,\vec{b})\in\shadowtable} \sum_{d}^{\nz} \invampspread{d} \Tr[U^\dagger \ketbra{\vec{b}}{\vec{b}} U \!\!\!\!\!\sum_{\substack{\opstring\text{ at}\\ \text{distance }d}} \!\!\!\! \opstring] .\label{eq:sumoverstringsimpl}
\end{align}
Naively, there are ${\volume \choose \nz}$ terms in this decomposition, each of which takes $\mathcal{O}(\nz)$ time to evaluate. 
However, the expectation values of uniformly delocalized $\{I, \Z\}$ strings depend only on the measurement string $\vec{b}$ and not the particular choice of 2-body gates $U_{[ij]}$. 
To compute \eqref{eq:sumoverstringsimpl}, we only need to read the measurement result in $\mathcal{O}(\volume)$ time, then perform a sum over $\mathcal{O}(4^{\nz})$ strings, calculating the trace for each.
The details of this algorithm are provided in Supp.~IIe, and pseudocode for implementing the fast expectation value calculation are provided in Supp.~IIf~\cite{supplemental}.

\paragraph{Sample Complexity ---}
The shadow-norm \eqref{eq:shadownorm} bounds the sample complexity of the All-Pairs channel.
For the same reasons which lead to the factorization in~\eqref{eq:channelsep}, the shadow norm can be separately evaluated for operator strings of $\plus$ and $\minus$, and strings of $I$ and $\Z$.
The latter is bounded using the eigenvalues $c_\lambda$. 
For $\volume \gg n_z \ge \lambda_2$, which is the relevant limit for reconstruction of low weight operators in large systems, the eigenvalues of $\mathcal{M}^{-1}$ scale as $c^{-1}_\lambda \simeq \left( \nicefrac{3}{2} \right)^{\lambda_2} - \mathcal{O}(\volume^{-1})$.
Accounting for the $a$, $a^\dagger$ pairs (Supp.~III~\cite{supplemental}) 
one obtains
\begin{align}
    \shadownorm{\opstring}^2 &\le \left(\frac{3}{2} \right)^{\!\!\np + 2\nz}  \!\left(\frac{\volume^{\np}}{\np!}\right) \norm{\opstring}_\infty^2 . \label{eq:shadownormsquared}
\end{align}
Notably, the shadow norm acquires a polynomial volume dependence compared to the single-unitary product channel without conserved charges, and retains a similar exponential dependence on the weight. 

\begin{figure}[t]
    \centering
    \includegraphics[width=1\linewidth]{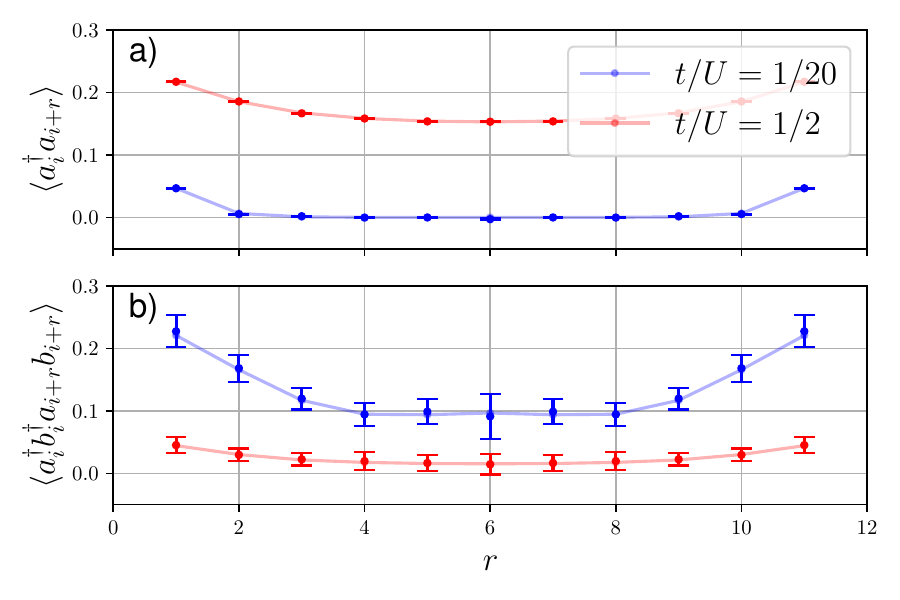}
    \vspace{-0.8cm}
    \caption{2- and 4-body correlation functions in a system of hardcore bosons on a two-leg ladder. The system realizes a paired/unpaired Luttinger liquid in the blue/red data. All-Pairs protocol estimates (markers) reconstructs correlation functions of the ground state (solid lines). We show the average and standard deviation of $50$ independent shadow tables with $N=2\times 10^4$ samples each.}
    \label{fig:laddercorrfuncs}
\end{figure}

\paragraph{Experimental considerations--- }

The essential experimental requirements to implement the All-Pairs protocol are: 
1. the ability to shuttle sites to bring paired sites into adjacency; 2. application of two-body gates to selected neighboring pairs;
and, 3. single-site occupation measurement.
Although our analysis above has assumed that the two-body gates are Haar random, it turns out that much simpler gate sets suffice.
For example, a uniform ensemble on the three gates
\begin{equation}
    \begin{aligned}
        G &= \{I, \sqrt{\text{iSWAP}}, \sqrt{\text{iSWAP}}\times (S \otimes I)\} 
    \end{aligned}
\end{equation}
suffices.
Here, $S$ is the  phase gate and $\sqrt{\text{iSWAP}}$ corresponds to a 50-50 beam splitter. 

\paragraph{Demonstration with paired Luttinger Liquid---} 
To demonstrate the utility of the All-Pairs channel, we illustrate its application to a system of hard-core bosons on a two-leg ladder which can be tuned to be a paired or unpaired Luttinger liquid. 
This example shows that the All-Pairs channel can be used to estimate arbitrary number-conserving correlation functions; it is \textit{not} limited to two-point functions.

We label the bosons $a$/$b$ on the top/bottom rails of a ladder with Hamiltonian $H = H_{\text{hop}}+H_{\text{int}}$, where
\begin{equation}
\begin{aligned}
    H_{\text{hop}}&=-t\sum_{i}\left[a_{i}^{\dagger}a_{i+1}+b_{i}^{\dagger}b_{i+1} + a_{i}^{\dagger}b_{i}\right]+\text{h.c.} \\
    H_{\text{int}} &= -U\sum_{i}a_{i}^{\dagger}a_{i}b_{i}^{\dagger}b_{i}  
\end{aligned}  
\end{equation}
The interaction term, $H_{\text{int}}$, is an attractive density-density interaction across rungs of the ladder. 

In the limit $t\gg U$, the bosons form a Luttinger liquid in which correlations of a single boson species decay as a power law, $\langle a^\dagger_i a_j\rangle \sim |i-j|^{-\alpha}$. When the interaction dominates, $t\ll U$, the bosons form `molecules' across rungs, which then condense into a paired liquid.
Here, the single-species correlators vanish exponentially, while pair correlations of the form $\langle a_{i}^{\dagger}b_{i}^{\dagger}a_{j}b_{j}\rangle$ exhibit power-laws.

We use the All-Pairs protocol to estimate single-boson and pair correlation functions in the ground state of $H$ obtained through exact diagonalization. 
We consider a ladder with 12 rungs ($\volume=24$ sites) at $1/4$ filling with periodic boundary conditions. 
Following Eq.~\eqref{eq:shadownormsquared}, we use $N \sim 2\times 10^4$ samples to get a bound on $\varepsilon \sim 1/5$ standard error on 4-body operators, although in practice the realized error is significantly smaller~\footnote{In a translationally invariant system, the sample complexity for the full correlation function does not differ significantly from that of a single operator. This can be understood as a special case of the double dixie cup problem~\cite{Newman_1960}.}. 
We construct 50 independent shadow tables to demonstrate that the empirical average estimator is unbiased and has small variance (cf. Fig~\ref{fig:laddercorrfuncs}). 
We note that the primary computational bottleneck in this exercise comes from acquiring the data, not the reconstruction itself; that is, most resources are consumed by exact diagonalization and sampling of the ground state.

\paragraph{Outlook. ---}
We have demonstrated that the All-Pairs protocol provides efficient local shadow tomography in systems with fundamental number conservation.
Thus, our protocol enables the tomographic investigation of strongly correlated many-body states in quantum simulators.
While we restricted to hard-core bosons in the main text, our results apply mutatis mutandis to fermions (Supp.~IIf~\cite{supplemental}).

Our algorithm for channel inversion allows for the estimation  of $\body$-body number-conserving correlation functions with resource requirements that scale favorably in comparison with alternative protocols (see Table~\ref{tab:channelsummary}).
Previous work has focused on Hamiltonian evolution~\cite{Mark_Choi_Shaw_Endres_Choi_2023,Tran_Mark_Ho_Choi_2023,Liu_Hao_Hu_2023}, which may be more practical to implement on some devices but requires an extremely accurate model of the underlying Hamiltonian in addition to significant classical post-processing resources.

Compared to systems without fundamental number conservation, the sample complexity has gained a volume dependence $\mathcal{O}(\volume^{\np})$.
We believe this volume scaling is optimal for estimating local observables.
In certain regimes shallow circuits can improve sample complexity in systems without number conservation~\cite{Akhtar_Hu_You_2023,Bertoni_Haferkamp_Hinsche_Ioannou_Eisert_Pashayan_2023,Ippoliti_Li_Rakovszky_Khemani_2023}, but it is unclear whether these improvements translate to systems with conservation laws.

There are two natural generalizations. 
Allowing internal states, the All-Pairs protocol generalizes by taking the two-body gates to be Haar on the local space. 
Allowing multiple occupancy ($M>1$ particles per site), one must extend to the ``All-Tuples'' protocol, in which the random pairings are replaced by random $(2M-1)$-tuples. 
In both of these cases, it is clear the channel remains tomographically complete and permutation symmetric.
We leave the analysis of the reconstruction algorithm and sample complexity to future work.

\begin{acknowledgments}
This work was supported by the Air Force Office of Scientific Research through grant No. FA9550-16-1-0334 (M.O.F.) and by the National Science Foundation through the awards DMR-1752759 (S.N.H., M.O.F. and A.C.) and PHY-1752727 (C.R.L.).
\end{acknowledgments}

\bibliography{bib}

\end{document}


\title{Efficient Local Classical Shadow Tomography with Number Conservation: Supplemental Material}

\author{Sumner N. Hearth}
\author{Michael O. Flynn}
\author{Anushya Chandran}
\author{Chris R. Laumann}
\affiliation{%
Department of Physics, Boston University, 590 Commonwealth Avenue, Boston, Massachusetts 02215, USA
}%

\date{\today}

\maketitle

\tableofcontents

\bgroup
\def\arraystretch{1.15}
\begin{table*}[!ht]
    \centering
    \begin{tabular}{c|c|c}
        Symbol & Description & Definition \\ \hline\hline
        $\rho$ & Quantum State &  \\
        $U_{[ij]}$ & Two-body number-conserving unitary & \\
        $\mathbf{b}$ & Measurement Result / bitstring & $\mathbf{b} = b_0 b_1 b_2 \ldots \in \{0, 1\}^{\otimes \volume}$ \\ \hline
        $[ij]$ & Two paired sites & $i \ne j$ \\
        $\pairing$ & A pairing configuration & $\pi = [ij][kl]\ldots \in \pairings{\volume}$ \\
        $\pairings{\volume}$ & The set of all pairings of $\volume$ sites & $\pairings{\volume} = \{\pairing_1, \pairing_2, \ldots, \pairing_{(\volume-1)!!}\}$ \\ 
        \hline
        $\plus$, $\minus$ & Bosonic creation and annihilation operators & $\plus = \begin{pmatrix}0 & 0 \\ 1 & 0\end{pmatrix}$ ~ $\minus = \begin{pmatrix}0 & 1 \\ 0 & 0\end{pmatrix}$  \\
        $\Z$ & The $Z$ operator for the 2-level system &  $Z = 1 - 2 \plus \minus$ \\
        $\nz$ & $\Z$ operator count in an operator string & \\
        $\np$/$\nm$ & $\plus$/$\minus$ operator count in an operator string & \\
        $\canopstring$ & A ``canonically ordered'' operator string & $\canopstring = \left(\plus \otimes \minus\right)^{\otimes \np} \otimes \Z^{\otimes \nz}$ \\
        \hline
        $\channel$ & The All-Pairs channel &  \\
        $\channel_{\pairing}$ & Channel action for pairing $\pi$ & Main text Eq.~(1)-(7) \\ 
        $\channel_{[ij]}$ & Channel action for two-site system &
        
    \end{tabular}
    \caption{Symbols and definitions for the main text}
\end{table*}
\egroup

\section{Number-Conserving Operators}

\subsection{Number-Conserving Operator Basis}
\label{app:operatorbasis}

In the main text, we introduced a local operator basis with single-body operators drawn from the set $\opset=\{I,\Z,\plus,\minus\}$.
Here we prove that the subset of number-conserving strings, denoted $\opset^{\otimes \volume}_N$, forms a basis for all number-conserving operators in a $\volume$-site system of hardcore bosons.

We begin by counting the dimension of the space of operators in this system. Let $\mathcal{H}$ denote the Hilbert space of states which can be partitioned into sectors with fixed boson number,
\begin{equation}    \mathcal{H}=\bigoplus_{N=0}^{\volume}\mathcal{H}_{N}
\end{equation}
The dimension of each number sector is given by a binomial coefficient, $\text{dim}(\mathcal{H}_{N}) = {\volume\choose N}$. The space of number-conserving operators on $\mathcal{H}$, denoted $\mathcal{LH}$, then has dimension
\begin{equation}\label{eq:OperatorCount1}
\begin{aligned}
    \text{dim}(\mathcal{LH}) &= \sum_{N=0}^{\volume}{\volume\choose N}^{2} = {2\volume\choose\volume}
\end{aligned}
\end{equation}

We will now compute the dimension of the space spanned by $\opset_{N}^{\otimes\volume}$. 
Operator strings $\opstring_{i},\opstring_{j}\in\opset_{N}^{\otimes\volume}$ are orthogonal with respect to the Hilbert-Schmidt norm:
\begin{equation}
\frac{1}{2^{\volume}}\Tr[\opstring_{i}^{\dagger}\opstring_{j}] \propto \delta_{ij}.
\end{equation}
Hence it is sufficient to count the number of strings in $\opset_{N}^{\otimes\volume}$.
Each element of $\opset_{N}^{\otimes\volume}$ contains equal numbers of $\plus$ and $\minus$ operators, so $\np\leq \left \lfloor{\volume/2}\right \rfloor $.
Once the $2\np$ locations of $\plus$ and $\minus$ are chosen, the remaining sites can be populated by $I$ and $\Z$ operators in $2^{\volume-2\np}$ ways.
Therefore the set of number-conserving operator strings has size
\begin{equation}\label{eq:OperatorCount2}
|\opset_{N}^{\otimes\volume}| = \sum_{\np=0}^{\left \lfloor{\volume/2}\right \rfloor}{\volume\choose 2\np}{2\np\choose \np}2^{\volume-2\np}
\end{equation}

It is not immediately obvious that equations~\eqref{eq:OperatorCount1} and~\eqref{eq:OperatorCount2} agree, but their equivalence can be established with generating functions. The following series representations of $(1-4x)^{-1/2}$ prove the result:
\begin{equation}
\begin{aligned}
    \frac{1}{\sqrt{1 - 4x}} &=\sum_\volume x^\volume \sum_{\np=0}^{\left \lfloor{\volume/2}\right \rfloor} {\volume \choose 2\np} {2\np \choose \np} 2^{\volume - 2\np}\\
    &=\sum_\volume x^\volume {2 \volume \choose \volume}
\end{aligned}
\end{equation}

\section{Diagonalizing and inverting $\channel$}\label{app:ChannelInversion}

In this appendix, we describe how to diagonalize and invert the All-Pairs channel restricted to the space spanned by $\{I,\Z\}^{\otimes V}$ operator strings.

\subsection{Overview}\label{sapp:Overview}

The channel $\channel$ has two key symmetry properties: it conserves the number $\nz$ of $\Z$ operators, and it is symmetric under permutations of $\volume$ sites $\sigma \in S_\volume$. 
This leads to two useful parameterizations of $\channel$.

First, we can expand the action of $\channel[\opstring]$ in terms of the swap distance $d$ to output strings, 
\begin{align}
    \channel[\opstring] &= \sum_{d=0}^{\nz} \ampspread{d} \sum_{\substack{\opstring_d \text{ at}\\ \text{distance }d}} \opstring_d . \label{eq:ampspreaddist}
\end{align}
Using simple combinatorial arguments, we will find a closed form for the amplitudes $\alpha_{d}$ in Supp.~\ref{sapp:StringSpreading}.

Second, we may block diagonalize $\channel$ into simultaneous irreducible representations (irreps) of $S_V$ and $\nz$. 
This corresponds to the usual decomposition of the action of the permutation group on $(\mathbb{C}^2)^{\otimes V}$~\cite{fulton1991representation}. 
\begin{align}
    \channel &= \sum_{\lambda} c_\lambda \sum_{n_{z}}\Pi_{\lambda,\nz} \label{eq:ProjectorDecomp}
\end{align}
Here, $\lambda = (V-\lambda_2, \lambda_2)$ represents a partition of $\volume$ into no more than two parts, as there are only two letters in the operator alphabet. 
The sum in Eq.~\eqref{eq:ProjectorDecomp} implicitly includes the constraint that $\lambda_2 \le \textrm{min}(\nz, V-\nz)$, as it is not possible to antisymmetrize pairs of identical characters.
We note two simplifying features: 1.  the irrep labeled $\lambda, \nz$ has multiplicity 1; and 2. it turns out that the eigenvalues $c_\lambda$ associated with $\Pi_{\lambda, \nz}$ is independent of $\nz$ (although this is not crucial for our analysis).

We can view the two decompositions Eq.~\eqref{eq:ampspreaddist} and \eqref{eq:ProjectorDecomp} as linear parametrizations of $\channel$ by the $\nz+1$ parameters $\ampspread{d}$ and $c_\lambda$, respectively, for each $\nz$. Thus, the parameters are linearly related,
\begin{align}
    \sum_{d} \ampspread{d} \ampeigmat_{\lambda d} = c_\lambda . 
    \label{eq:linearrel}
\end{align}
In Supp.~\ref{sapp:ChannelEigenvalues}, we find a closed-form expression for the $(\nz+1) \times (\nz + 1)$ integer matrix $\ampeigmat_{\lambda d}$ which, in turn, furnishes closed-form expressions for the eigenvalues $c_{\lambda}$ building on our knowledge of $\ampspread{d}$ from Supp.~\ref{sapp:StringSpreading}.

The inverse channel $\channel^{-1}$ is particularly simple in the irrep decomposition,
\begin{align}
    \channel^{-1} &= \sum_{\lambda} \frac{1}{c_\lambda} \sum_{n_{z}}\Pi_{\lambda,\nz} \label{eq:InvProjectorDecomp}
\end{align}
In order to efficiently compute operator estimators, the form in terms of swap distance is more useful,
\begin{align}
    \channel^{-1}[\opstring] &= \sum_{d=0}^{\nz} \invampspread{d} \sum_{\substack{\opstring_d\text{ at}\\ \text{distance }d}} \opstring_d \label{eq:invampspreaddist}
\end{align}
where, the $G$ matrix relates the amplitudes $\beta_d$ to the eigenvalues of $\channel$,
\begin{align}
    \sum_d \invampspread{d} \ampeigmat_{\lambda d} &= \frac{1}{c_\lambda} \label{eq:invlinearmap}.
\end{align}
We do not have a closed form for $G^{-1}$ but it can be precomputed efficiently.

With the amplitudes $\invampspread{d}$ in hand, it is now possible to compute estimators using the data in a shadow table $T$,
\begin{align}
    \Tr[\rho \opstring] &= \frac{1}{|T|} \sum_{(U,\vec{b})\in\shadowtable}\Tr[\left(U^\dagger \ketbra{\vec{b}}{\vec{b}} U\right) \channel^{-1}[\opstring]] \label{eq:estimatorbasic} \\
    &= \frac{1}{|T|} \sum_{(U,\vec{b})\in\shadowtable} \sum_{d} \invampspread{d} \Tr[\left(U^\dagger \ketbra{\vec{b}}{\vec{b}} U\right) \!\!\!\!\!\!\sum_{\substack{\opstring_d\text{ at}\\ \text{distance }d}} \!\!\!\! \opstring_d] \label{eq:sumoverstringsimpl}
\end{align}
In Supp.~\ref{sapp:Implementation}, we develop an efficient strategy for evaluating Eq.~\eqref{eq:sumoverstringsimpl}. Although this computation appears to be a sum over ${\volume \choose \nz}$ terms, the expectation value can be computed in only $\mathcal{O}(\volume + \nz 4^{\nz})$ time by taking advantage of the fact that $\nz\ll\volume$ in a typical application.

\subsection{Computing Swap Amplitudes $\ampspread{d}$}
\label{sapp:StringSpreading}

The amplitude $\ampspread{d}$ depends on the volume $\volume$ and the $\nz$-sector. 
Our recursive construction of $\ampspread{d}$ requires keeping track of these dependencies so we write $\ampspreadfull{d}{\volume}{\nz}$.

The base case for the recursion is the amplitude $\alpha_{0}^{\volume,\nz}$ which remains on an input string $\opstring$ following the action of the channel,
\begin{align}
    \ampspreadfull{0}{\volume}{\nz} = \frac{1}{2^\volume} \Tr[\opstring^\dagger \channel[\opstring]] .
\end{align}
In the space of $\{I,\Z\}$ strings, the superoperator $\channel_{[ij]}$ acts as:
\begin{equation}
\channel_{[ij]}\left[\opstring_i\opstring'_j\right] = 
\begin{cases}
    \opstring_i \opstring'_j, & \opstring=\opstring'\\
        \frac{2}{3}\opstring_i\opstring'_j + \frac{1}{3}\opstring'_j\opstring_i, & \opstring\neq\opstring'
\end{cases}
\end{equation}
The swap distance increases only if the operators on sites $i$ and $j$ are distinct and does so with probability $1/3$. 
The amplitude $\ampspreadfull{0}{\volume}{\nz}$ follows from counting the number of pairings which link $m$ pairs of $(I,\Z)$ operators:
\begin{widetext}
\begin{align}\label{eq:Zeroswapweight}
    \ampspreadfull{0}{\volume}{\nz} &= \frac{1}{| \pairings{\volume} |} \sum_{\substack{m=0\\m \cong \nz \text{ mod }2}}^{\nz} \left(\frac{2}{3}\right)^m \underbrace{{\nz \choose m}{{\volume - \nz} \choose m} m!}_{\text{Number of $I \Z$ pairings}} \times  \underbrace{|\pairings{\nz - m}| |\pairings{\volume - \nz - m}|}_{\substack{\text{Number of $\Z\Z$}\\\text{and $II$ pairings}}} .
\end{align}
\end{widetext}
We recall that $|\pairings{\volume}| = (\volume-1)!!$ denotes the number of pairings of $\volume$ objects.
As a simple consistency check, we note that the strings consisting entirely of $I$'s or $\Z$'s must be eigenoperators of $\channel$ due to conservation of $\nz$. This is reflected in Eq.~\eqref{eq:Zeroswapweight}: $\ampspreadfull{0}{\volume    }{0}=\ampspreadfull{0}{\volume}{\volume}=1$.

The general amplitude function $\ampspreadfull{d}{\volume}{\nz}$ then obeys
\begin{align}
    \alpha_d^{V, n_z} = \frac{1}{3^d}
    \times \frac{d! |P_{V-2d}|}{|P_V|} \times  \alpha_0^{V-2d, n_z - d} \label{eq:ampspreadrec}
\end{align}
This equation relates the amplitude to reach a string at swap distance $d$ to the amplitude to leave $V-2d$ letters invariant. 
The first factor is the reduction of amplitude for $d$ non-trivial swaps while the second is the fraction of pairings which pair amongst the $2d$ swapped letters.

\subsection{Computing $G$ and Eigenvalues $c_\lambda$}\label{sapp:ChannelEigenvalues}

To compute the eigenvalues $c_\lambda$ and linear relations $\ampeigmat_{\lambda d}$, we use a representative operator in $\Pi_{\lambda, n_z}$. 
Let us set up some formalism.

We call an operator string canonically ordered if all $Z$ operators precede $I$ operators. We denote such operators by $\canopstring=\Z^{\otimes\nz}\otimes I^{\otimes\volume-\nz}$.

\begin{figure}[ht]
    \centering
    \includegraphics[width=0.75\columnwidth]{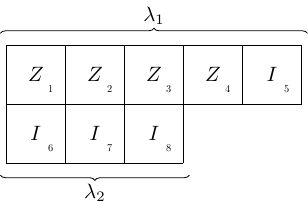}
    \caption{An example eigenoperator represented by a decorated Young diagram with up to two rows. Each vertical column represents an antisymmetric pair of indices, rows represent the sets of symmetric indices.}
    \label{fig:youngtableauexample}
\end{figure}

We can find a representative operator in $\Pi_{\lambda, n_z}$ by using a Young symmetrizer $\Pi_T$ on the reference string $\canopstring$ for a given irrep $\lambda$ of $S_V$. 
Take $T$ to be a canonically ordered Young tableau corresponding to the Young diagram $\lambda$, see Fig.~\ref{fig:youngtableauexample}.
The Young symmetrizer $\Pi_T$ acts on operator strings by symmetrizing indices within rows and anti-symmetrizing indices within columns, 
\begin{align}
    \Pi_T &= \left(\prod_{\langle i j \rangle } \left(\mathbbm{1}_{ij} - \mathbb{S}_{ij}\right)\right)\left(\sum_{\substack{\sigma_{\text{top}}\in S_{\lambda_{1}} \\ \sigma_{\text{bot}} \in S_{\lambda_2}}} \sigma_{\text{top}} \otimes \sigma_{\text{bot}} \right) .
\end{align}
with $\mathbbm{1}_{ij}$ an identity superoperator, $\mathbb{S}_{ij}$ a swap, and $\sigma$ a permutation of the sites within each row.

As a concrete example, consider an operator string with $\volume=8$ sites and the irrep described by the diagram in Fig.~\ref{fig:youngtableauexample}.
The figure shows both a Young tableau with $\lambda_2 = 3$ antisymmetric index pairs ($\langle 1, 6\rangle$, $\langle 2, 7\rangle$, and $\langle 3, 8\rangle$) and an input string $\canopstring = \Z_1 \Z_2 \Z_3 \Z_4$.
The Young-symmetrized output, $\Pi_T\left[\canopstring\right]$, is given by
\begin{equation}
\begin{aligned}
    \op_{T} &= \Pi_T\left[\canopstring\right] \\&= (\Z_1 - \Z_6)(\Z_2 - \Z_7)(\Z_3 - \Z_8)(\Z_4 + \Z_5) . \label{eq:exampleeigen}
\end{aligned}
\end{equation}
The output string is clearly antisymmetric under swaps of indices $\langle i j \rangle$ ($\mathbb{S}_{ij}[\op_T] = - \op_T$), and symmetric under permutations of indices $\{4, 5\}$. 

To compute the eigenvalues, we exploit 
\begin{align}\label{eq:eigenvalueamp}
    c_\lambda = \frac{\Tr[\canopstring \channel[ \Pi_T \canopstring]]}{\Tr[\canopstring \Pi_T \canopstring]} = \frac{1}{2^\volume} \Tr[\canopstring \channel[\Pi_T \canopstring]]
\end{align}
Each string $\opstring_d$ in the support of $\Pi_T \canopstring$ comes with a sign determined by the particular Young tableau $T$ and is at some swap-distance $d$ from $\canopstring$. 
In order to contribute to Eq.~\eqref{eq:eigenvalueamp}, the channel $\channel$ acting on $\opstring_d$ must return to $\canopstring$, that is $\channel$ should decrease the swap-distance by $d$.
Thus, the net amplitude which returns to $\canopstring$ is just the (signed) sum of contributions from strings over all values of $d$,
\begin{align}
    c_\lambda &= \sum_{d=0}^{\nz} \ampspread{d} \ampeigmat_{\lambda d} \label{eq:eigenvaluesecondtime}\\
    \ampeigmat_{\lambda d} &= \sum_{x+y = d} \underbrace{\left(-1\right)^x {\lambda_2 \choose x}}_{\text{antisymmetric}} \underbrace{{{\nz-\lambda_2} \choose y} {{\volume-\nz-\lambda_2} \choose y}}_{\text{symmetric}} \label{eq:amptoeigmatrix}.
\end{align}
Here $x$ is the number of antisymmetric swaps picked from $\lambda_2$ possibilities and $y = d - x$ is the number of symmetric swaps between the remaining $\Z$ and $I$ operators. 
\emph{En passant}, we have obtained both $\ampeigmat$ and the eigenvalues $c_\lambda$. 

It is not obvious from Eq.~\eqref{eq:eigenvaluesecondtime} that the eigenvalues are independent of $\nz$. However, the dynamics of $\{I,Z\}$ strings are equivalent to the dynamics of hardcore random walkers executing a symmetric exclusion process, which is known to be non-interacting; we therefore expect the eigenvalues to be $\nz$-independent. 
Numerics confirm this. 
We are thus free to evaluate $c_{\lambda}$ at convenient values of $\nz$; setting $\nz=\lambda_2$, we find that
\begin{widetext}
    \begin{align}
        c_\lambda &= \sum_{d=0}^{\lambda_2} \sum^{\lambda_2 - d}_{\substack{m=0\\m\cong \lambda_2 - d \text{ mod } 2}} \left(-\frac{1}{3}\right)^d \left(\frac{2}{3}\right)^m \frac{\lambda_2!}{ \left(\lambda_2 - d - m\right)!} {\volume - d - \lambda_2 \choose m}\left(\frac{|\pairings{\lambda_2 - d - m}| |\pairings{\volume - d - \lambda_2 - m}|}{|\pairings{\volume}|}\right) \label{eq:eigenfullconj}
    \end{align}
\end{widetext}

In the large $\volume$ limit, the volume dependence in the summand is simple:
\begin{align}
    \lim_{\volume \rightarrow \infty} {{\volume - d - \lambda_2} \choose m}\frac{|\pairings{\volume-d-\lambda_2 - m}|}{|\pairings{\volume}|} \propto \volume^{-\frac{1}{2}(d - m + \lambda_2)}
\end{align}

\begin{figure}
    \centering
    \includegraphics[width=\linewidth]{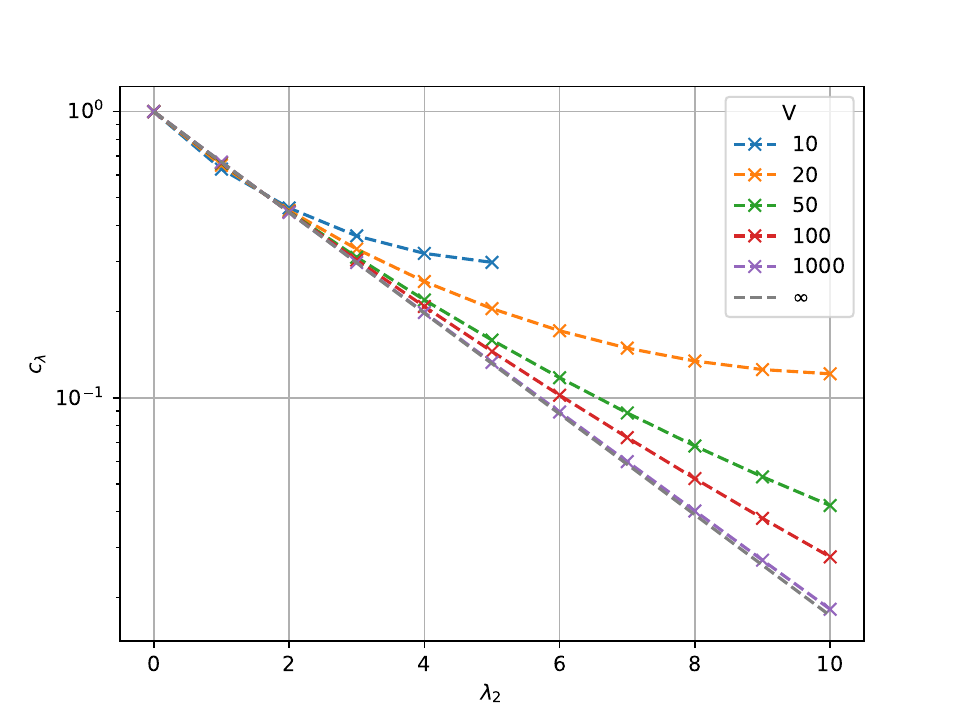}
    \caption{Eigenvalues of $\channel$ for different symmetry sectors labeled by partition $\lambda=(\lambda_1, \lambda_2)$. The eigenvalues only depend on $\lambda_2$ and $\volume = \lambda_1 + \lambda_2$.}
    \label{fig:finitesizedeigs}
\end{figure}

The exponent of $\volume$ is maximized by $m=\lambda_2 - d$, $d=0$, which yields a leading order contribution that is volume-independent. 
Thus, the largest contribution to $c_{\lambda}$ comes from zero-swap distance,
\begin{align}
    \lim_{\volume \rightarrow \infty} c_\lambda &\simeq \left(\frac{2}{3}\right)^{\lambda_2} + \mathcal{O}\left(\frac{1}{\volume}\right) . \label{eq:eigenlargevapprox}
\end{align}
Intuitively, larger swap distances are suppressed because the string amplitude is distributed among the $\mathcal{O}({\volume \choose d})$ strings at fixed swap distance.
In Fig.~\ref{fig:finitesizedeigs} we show the convergence of \eqref{eq:eigenfullconj} to \eqref{eq:eigenlargevapprox} with increasing $\volume$.

\subsection{Implementation Details}\label{sapp:Implementation}

We are now prepared to estimate expectation values of observables $\opstring$.
We focus on calculating $\bra{\vec{b}} U \channel^{-1}[ \canopstring ] U^\dagger \ket{\vec{b}}$ with $\canopstring = \Z^{\otimes \nz} \otimes I^{\otimes \volume - \nz}$.

Using the swap distance representation of $\channel^{-1}$, we have
\begin{align}
    \bra{\vec{b}} U \channel^{-1}[ \canopstring ] U^\dagger \ket{\vec{b}} &= \sum_{d=0}^{\nz} \invampspread{d} \!\!\!\! \sum_{\substack{\opstring_d \text{ at} \\ \text{distance }d}} \!\!\!\! \bra{\vec{b}} U \opstring_d U^\dagger \ket{\vec{b}} . \label{eq:delocalizedzclean} 
\end{align}
We compute the amplitudes $\invampspread{d}$ numerically via the relation~\eqref{eq:invlinearmap} by inverting the $(\nz+1) \times (\nz+1)$ integer matrix $G$~\eqref{eq:amptoeigmatrix} and applying it to the vector of (inverse) eigenvalues $1/c_\lambda$~\eqref{eq:eigenfullconj}.

There are ${\volume \choose \nz}$ terms in the sum over strings \eqref{eq:delocalizedzclean}, so we  na\"ively expect a computational overhead which grows as $\mathcal{O}(\volume^{\nz})$. 
This computational cost can be significantly reduced in the $\nz \ll \volume$ limit by factoring~\eqref{eq:delocalizedzclean} into a sum over strings in two spatial regions, one of which is highly constrained.
Given a pairing $\pi\in\pairings{\volume}$, recall that each unitary in the All-Pairs protocol is the product of $\volume/2$ two-body unitaries,
\begin{align}
    U = \bigotimes_{[ij] \in \pi} U_{[ij]} .
\end{align}
It is convenient to use notation which makes operator pairings manifest. 
We write each operator string in a two-row format such that each column represents two sites paired by a unitary. 
For example, consider a four-site subsystem where unitaries pair sites $[13]$ and $[24]$. 
Then we choose to rearrange sites to obtain the representation
\begin{align}
    \begin{tikzpicture}[scale=0.75,baseline={([yshift=-.5ex]current bounding box.center)}]
        \node at (-1,0) {$\canopstring = $};
        \foreach \x in {0,...,2} {
            \node at (0.5*\x,0) {Z};
        }
        \foreach \x in {3,...,3} {
            \node at (0.5*\x,0) {I};
        }
        \node at (2,0) {$\ldots$};
        \draw[<->] (0,0.25) to[out=45,in=135] (1,0.25);
        \node at (0.5,0.75) {$U_{[13]}$};
        \draw[<->] (0.5,-0.25) to[out=-45,in=-135] (1.5,-0.25);
        \node at (1,-0.75) {$U_{[24]}$};
    \end{tikzpicture} =
    \begin{array}{ccc}
        \Z_1 & \Z_2 & \ldots \\
        \Z_3 & I_4 & \ldots
    \end{array}
\end{align}
The following arguments do not depend on the specific site index in each column so we drop subscripts.

As a reference example, consider a string of size $\volume=16$ and $\nz=6$,
\begin{align}
    \canopstring &= \underbrace{\begin{array}{cccc|}
        \Z & \Z & \Z & \Z \\
        \Z & \Z & I & I
    \end{array}}_{\text{region }A}\underbrace{\begin{array}{cccc}
        I & I & I & I \\
        I & I & I & I
    \end{array}}_{\text{region }B} \label{eq:examplerewrite}.
\end{align}
A vertical line separates the sites into two regions, $A$ and $B$.
Region $A$ contains columns with non-identity operators in at least one site, while region $B$ contains only $I-I$ pairings.

To generate a string $\opstring_d$ at swap-distance $d$ from $\canopstring$, we swap $d$ of the $\Z$ operators with $I$'s, in either region $A$ or $B$. For example, starting from the reference string of Eq.~\eqref{eq:examplerewrite}, a valid $d=2$ string is
\begin{align}\label{eq:exampledisttwo}
    \opstring_{d=2} &= \begin{array}{cccc|cccc}
        \Z & \Z & {\color{red}I} & \Z & I & I & I & {\color{red}\Z} \\
        {\color{red}I} & \Z & {\color{red}\Z} & I & I & I & I & I
    \end{array}
\end{align}
where the swapped operators are shown in red.

Given the string in Eq.~\eqref{eq:exampledisttwo}, any permutation of the sites in region $B$ leaves the swap-distance relative to $\canopstring$ unchanged.  
Thus in Eq.~\eqref{eq:delocalizedzclean}, we can treat the sum over region $B$ as though it is in a uniform superposition of all strings consistent with the state of $A$.

Let us now formalize this argument. Let $\mathcal{S}_{k_{b}}$ denote the set of all strings in region $B$ which contain $k_{b}\; \Z$ operators. Let $\mathcal{S}_{d,k_a}$ denote the set of all strings in region $A$ constructed by changing $d$ of the $\Z$ operators to identities, and $k_a$ identities to $\Z$ operators.
We can then rewrite the sum over all strings at swap-distance $d$ as a sum over strings from each of these sets,
\begin{align}
    \sum_{\opstring_d} \opstring_d = \sum_{k_a+k_b = d} \left(\sum_{\opstring_A \in \mathcal{S}_{d,k_a}} \opstring_A \right) \otimes \left(\sum_{\opstring_B \in \mathcal{S}_{k_b}}  \opstring_B \right) \label{eq:computebreakdown}
\end{align}
Expectation values similarly decompose into the product of the expectation values on regions $A$ and $B$,
\begin{widetext}
\begin{align}
    \bra{\vec{b}}\! U \channel^{-1}[ \canopstring ]& U^\dagger \! \ket{\vec{b}} = \sum_d \invampspread{d} \sum_{k_a+k_b = d} \bra{b_A} U_A \left( \sum_{\opstring_A \in \mathcal{S}_{d,k_a}} \opstring_A \right) U^\dagger_A \ket{b_A} \times \bra{b_B} U_B \left(\sum_{\opstring_B \in \mathcal{S}_{k_b}} \opstring_B\right) U^\dagger_B \ket{b_B} \\
    &= \sum_d \invampspread{d} \!\! \sum_{k_a+k_b = d} \! \left(\sum_{\opstring \in \mathcal{S}_{d,k_a}} \prod_{[ij] \in \pairing_A} \!\!\! \bra{b_i b_j} U_{[ij]} \opstring_i \opstring_j U^\dagger_{[ij]} \ket{b_i b_j} \right) \!\! \underbrace{\left(\sum_{\opstring \in \mathcal{S}_{k_b}} \prod_{[ij] \in \pairing_B} \!\!\! \bra{b_i b_j} U_{[ij]} \opstring_i \opstring_j U^\dagger_{[ij]} \ket{b_i b_j} \right)}_{h(\pairing_B, \vec{b}, k_b)} \label{eq:delocalizedzcalc}
\end{align}
\end{widetext}
where we have denoted the contribution due to the uniform sum over allowed strings in region $B$ by $h(\pairing_B, \vec{b}, k_b)$.

The term $h(\pairing_B, \vec{b}_B, k_b)$ can be computed in a time which scales as $\mathcal{O}(\volume + k_b^3)$.
Region $B$ is initially populated only by identity operators,
\begin{align}
    \begin{array}{cccc}
        I & I & I & I \\
        I & I & I & I
    \end{array} .
\end{align}
To construct a string $\opstring$ with $k_b$ $\Z$ operators, choose $k_b$ of the identity operators to replace with $\Z$'s.
These $\Z$ operators can either be unpaired, meaning that they occupy a column with an $I$, or be paired with another $\Z$.
It is helpful to organize the construction by first choosing how many $\Z$s will be paired, which we call $m$, and the positions of these pairs.
We then place the remaining $k_b - 2m$ unpaired $\Z$s among the remaining columns; for example,
\begin{align}
    \begin{array}{cccc}
        I & I & I & I \\
        I & I & I & I
    \end{array} \rightarrow
    \begin{array}{cccc}
        I & {\color{red}\Z} & I & I \\
        I & {\color{red}\Z} & I & I
    \end{array}
    \rightarrow
    \begin{array}{cccc}
        {\color{red}\Z} & \Z & I & I \\
        I & \Z & I & {\color{red}\Z}
    \end{array}
\end{align}
Consider a column $[ij]$ which contains an unpaired $\Z$. This $\Z$ could be placed in either site $i$ or $j$, so the uniform sum over all $\opstring \in \mathcal{S}_{k_b}$ can be regrouped into superpositions $\Z_i I_j + I_i \Z_j$,
\begin{align}
    \begin{array}{cccc}
        I & {\color{red}\Z} & I & I \\
        I & I & I & I
    \end{array} + 
    \begin{array}{cccc}
        I & I & I & I \\
        I & {\color{red}\Z} & I & I
    \end{array} =
    \begin{array}{c}
        I \\
        I 
    \end{array}\left(\begin{array}{c}
        {\color{red}\Z} \\
        I 
    \end{array} + \begin{array}{c}
        I \\
        {\color{red}\Z} 
    \end{array}\right)
    \begin{array}{cc}
        I & I \\
        I & I
    \end{array}
\end{align}
This has the benefit of making all three possible column configurations ($I_i I_j$, $Z_i Z_j$, and $\Z_i I_j + I_i \Z_j$) commute with all number-conserving operators.
Expectation values can then be expressed solely in terms of the local measurement results $b_i$ and $b_j$,

\begin{equation}
\begin{aligned}
    \bra{b_i b_j} U_{[ij]} \left(I_i I_j\right) U_{[ij]}^\dagger \ket{b_i b_j} &= 1 \\
    \bra{b_i b_j} U_{[ij]} \left(\Z_i \Z_j\right) U_{[ij]}^\dagger \ket{b_i b_j} &= (-1)^{b_i + b_j} \\
    \bra{b_i b_j} \! U_{[ij]} (\Z_i I_j + I_i \Z_j) U_{[ij]}^\dagger \! \ket{b_i b_j} &= 2(-1)^{b_i} \delta_{b_i b_j} .
\end{aligned}
\label{eq:zzfactors}
\end{equation}
As there are only four possible local measurement results (00, 01, 10, 11), the entire contribution from region $B$ is a combinatorial expression determined by the number of ways to allocate $\Z$ operators between the four types of columns.
Let $N_{bb'}$ be the number of columns with measurement result $bb'$,
\begin{align}
    N_{bb'} &= \sum_{[ij] \in \pairing_B} \delta[bb' = b_i b_j] .
\end{align}
We now count the number of ways to place $m$ pairs ($\Z_i \Z_j$) and the remaining $k_b - 2m$ superpositions ($\Z_i I_j + I_i \Z_j$).
Each placement into a column with local measurement result $bb'$ is associated with an amplitude given by Eq.~\eqref{eq:zzfactors}.
For example, if a pair $\Z_i \Z_j$ is placed into a column with measurement result $b_i b_j$ it will contribute a factor of $(-1)^{b_i + b_j}$.
The net contribution from region $B$ can then be calculated by summing over all possible distributions over columns,
\begin{widetext}    
\begin{align}
    h(\pairing_B,\vec{b},k_b) &\equiv \sum_{\opstring \in \mathcal{S}_{k}} \prod_{[ij] \in \pairing_B} \bra{b_i b_j} U_{[ij]} \opstring_i \opstring_j U^\dagger_{[ij]} \ket{b_i b_j} \\
    &= \sum^{k_b/2}_m 
    \underbrace{\left(\sum_{x+y=m} (-1)^y {{N_{00} + N_{11} - (k_b - 2m)} \choose x}{{N_{01} + N_{10} \choose y}} \right)}_{\text{place $m$ pairs $\Z\Z$}}
    \underbrace{\left(\sum_{x+y=k_b-2m} 2^{x+y} (-1)^y{N_{00} \choose x}{N_{11} \choose y}\right)}_{\text{place $k_b-2m$ superpositions $\Z I + I \Z$}}
      \label{eq:efficientdeloc} 
\end{align}
\end{widetext}

The function $h(\pairing_B,\vec{b},k_b)$ may be computed once per pairing $\pairing$ for each value of $k_{b}$ ($0 \le k_b \le \nz$) with a net post-processing time of $\mathcal{O}(\volume + \nz^{\!\!3})$.
The leading contribution to the computational complexity then comes from the sum over strings in region $A$. In the worst case, region $A$ contains $\nz$ $\Z$ operators and $\nz$ identities, and the number of strings grows as
\begin{align}
    \sum_{d=0}^\nz  \sum_{k_a=0}^d  |\mathcal{S}_{d,k_a}| = \sum_{d=0}^\nz  \sum_{k_a=0}^d {\nz \choose d} {\nz \choose k_a} < 4^\nz .
\end{align}
For each string, computing the trace takes $\mathcal{O}(\nz)$ time since both $U$ and $\opstring$ are product operators.
The final runtime complexity is then bounded by $\mathcal{O}(\volume + \nz 4^\nz)$ per sample, plus a one-time cost of $\mathcal{O}(\nz^{\!\!3})$ from inverting $\ampeigmat$.

\subsection{Pseudocode}\label{sapp:Pseudo}

Here we present pseudocode for efficiently computing the expectation value of $\canopstring$ using $h$ as defined in Eq.~\eqref{eq:efficientdeloc}.
The function \verb|S(So, d,ka)| represents the set $\mathcal{S}_{d,k_a}$ used in Eq.~\eqref{eq:computebreakdown}, and the \verb|expect| function represents the expectation value taken over strings in region $A$.
A full implementation is available online~\cite{Hearth_github_2023}.

\begin{widetext}
\begin{verbatim}
    compute g_inverse;
    compute c;
    beta = g_inverse @ (1/c); 

    fn compute_sample(So, pi, Us, b)
        pi_A = {(i,j) in pi : i<nz or j<nz};
        pi_B = pi - pi_A;
        
        precompute h(pi_B, b[p_B], 0...nz);

        total = 0;
        for d in 0...nz
            for ka in 0...d
                for s in S(So, d, ka)
                    total += beta[d] * h[d-ka] * expect(pi_A, Us, b, s);
                end
            end
        end
        return total;
    end
\end{verbatim}
\end{widetext}

\subsection{Spinless Fermions}

Adjusting the All-Pairs protocol to the case of spinless fermions is straightforward.
We construct spinless fermion operator strings $S \in \{I, \Z, c^\dagger, c\}^{\otimes \volume}$ using fermionic creation and annihilation operators.
The action of $\channel$ on these operator strings is identical to the action on bosonic operators, and the same reconstruction algorithm described from hardcore bosons also applies to spinless fermions.
To see this, we first note that the two-site channel $\channel_{[ij]}$ acting on adjacent sites $i$ and $j$ is identical to the bosonic case,

\bgroup
\begin{table}[h!]
    \centering
    \def\arraystretch{1.2}
    \begin{tabular}{c|c}
        $\opstring_i \opstring_j$ & $\channel_{[ij]}[\opstring_i \opstring_j]$\\ \hline\hline
        $I_i I_j$ & $I_iI_j$ \\
        $\Z_i\Z_j$ & $\Z_i\Z_j$ \\\hline
        $I_i\Z_j$ & $\frac{2}{3}I_i\Z_j + \frac{1}{3}\Z_iI_j$ \\
        $\Z_iI_j$ & $\frac{2}{3}\Z_iI_j + \frac{1}{3}I_i\Z_j$ \\\hline
        $c^\dagger_i c_j$ & $\frac{1}{3}c^\dagger_i c_j$ \\
        $c_i c^\dagger_j$ & $\frac{1}{3}c_i c^\dagger_j$ \\\hline
        \text{Else} & 0
    \end{tabular}
\end{table}
\egroup

Second, we explicitly write the channel in terms of permutation $\sigma$ and $\channel_{\mathbbm{1}} = \bigotimes_{i}^{\volume/2} \channel_{[2i,2i+1]}$,
\begin{align}
    \channel[\rho] = \frac{1}{|\volume!|} \sum_{\sigma} \sigma^\dagger \channel_{\mathbbm{1}}[\sigma \rho \sigma^\dagger] \sigma . \label{eq:explicitrewrite}
\end{align}
Each permutation $\sigma$ takes sites $\sigma^{-1}(2i)$ and $\sigma^{-1}(2i+1)$ to adjacent sites $2i$ and $2i+1$, which are then acted on by the 2-body channel $\channel_{[2i,2i+1]}$.
Since unpaired $c^\dagger$ and $c$ operators are annihilated by the 2-body channel, $\channel$ once again factorizes into the action on creation/annihilation operators and the action on $I$ and $\Z$ operators.
The action of $\channel$ on a fermionic operator string is therefore identical to the case of bosonic operators.

Consider the expectation value of the canonically ordered operator string $\canopstring$,
\begin{equation}\begin{aligned}
    \langle \canopstring \rangle &= \frac{1}{|\shadowtable|} \sum_{(\sigma, U, \mathbf{b})\in\shadowtable} \bra{\mathbf{b}} \sigma^\dagger U \sigma \channel^{-1}[\canopstring] \sigma^\dagger U^\dagger \sigma \ket{\mathbf{b}}  \\
    &= \frac{1}{|\shadowtable|} \sum_{(\sigma, U, \mathbf{b})\in\shadowtable} \bra{\sigma(\mathbf{b})} U \sigma  \channel^{-1}[\canopstring] \sigma^\dagger U^\dagger \ket{\sigma(\mathbf{b})} .\label{eq:fermionicreconstruction}
\end{aligned}\end{equation}
As before, only permutations which do not mix creation/annhilation operators with $\Z$ and $I$ operations may contribute; for each term in the sum we may rewrite the expectation value as the product of the expectation value of $S_A = \left(c^\dagger \otimes c\right)^{\otimes \np}$ and $S_B = \Z^{\otimes \nz}$.
The evaluation of $S_B$ proceeds exactly as in the bosonic case described in \ref{sapp:Implementation}, and the evaluation of $S_A$ is straightforward using standard second quantized sign rules.

\section{Shadow norm evaluation for the All-Pairs protocol}
\label{app:shadownorm}

The sample complexity of estimating an operator string $\opstring$ is controlled by the variance of the contributions from individual samples.
This variance is bounded by the shadow norm $\shadownorm{\opstring}^2$.
In this appendix, we calculate the shadow norm by factorizing the contributions into those from $\plus$ and $\minus$ operators and those from $I$ and $\Z$ operators.
To begin, recall that the shadow norm of an operator $\opstring$ and channel $\channel$ is given by
\begin{align}
    \shadownorm{\opstring}^2 = \max\limits_{\rho}\! \mathop{\mathbb{E}}\limits_{\pairing, U_{\pairing}} \sum_b \bra{\vec{b}} U_{\pairing} \rho U_{\pairing}^\dagger \ket{\vec{b}} \left|\bra{\vec{b}} U_{\pairing} \channel^{-1}[\opstring] U_{\pairing}^\dagger \ket{\vec{b}}\right\vert^2 \!\!.
\end{align}

Given a string $\opstring$, define region $A$ to consist of sites with $\plus$ and $\minus$ operators, and region $B$ to consist of sites hosting $I$ and $\Z$ operators. Contributions to the shadow norm come from pairings which link each $\plus$ with a $\minus$; therefore, the set of relevant pairings do not link any sites between regions $A$ and $B$.
The number of such pairings is given by $f |\pairings{\volume}|$, where $f$ is defined in (11) in the main text.
Correspondingly, we find that the state $\rho^\star$ which maximizes the shadow norm is that which independently maximizes the contributions of region $A$ and region $B$,
\begin{align}
\begin{aligned}
    \rho^\star &= \rho^\star_A \otimes \rho^\star_B
\end{aligned}
\end{align}
and the shadow norm factorizes,
\begin{align}
    \shadownorm{\opstring }^2 = \shadownorm{ \opstring_A }^2 \times \shadownorm{ \opstring_B }^2 .
\end{align}

Consider first the shadow norm of region $A$.
We bound the maximization over $\rho^\star_A$ by an independent maximization over $\rho_{[ij]}$ for each pair of sites in $\pairing$,
\begin{align}
    \shadownorm{\opstring_A }^2 &= \frac{3^{2\np}}{f^2} \max\limits_{\rho_A} \!\! \mathop{\mathbb{E}}\limits_{\pairing, U_{\pairing}, \vec{b}} \vert\!\bra{\vec{b}} U_{\pairing}\!\left(\plus\!\otimes\!\minus\right)^{\!\!\oplus\np} U_{\pairing}^\dagger \ket{\vec{b}}\!\vert^2 \\
    &\le \frac{1}{f} \Lambda^{\np} \\
    \Lambda &= \max_{\rho} \mathop{\mathbb{E}}\limits_{U b b'}\left| \bra{b b'} U \left(3 \plus \otimes \minus\right) U^\dagger \ket{b b'} \right|^2 \\
    &= \frac{3}{2} \max_{\rho} \left[ \bra{01}\rho\ket{10} + \bra{10}\rho\ket{01} \right] = \frac{3}{2}
\end{align}

Next, consider region $B$.
Among these sites, $\Z$ operators are delocalized in accordance with Eq.~\eqref{eq:delocalizedzclean}.
To bound the shadow norm, we note that an operator can be rescaled by no more than the largest eigenvalue of $\channel^{-1}$,
\begin{align}
    \shadownorm{\opstring_B}^2 &= \max_{\rho_B} \mathop{\mathbb{E}}\limits_{\pairing,U_{\pairing},\vec{b}} \bra{\vec{b}} U_\pairing \channel^{-1}[\Z^{\otimes \nz}] U_\pairing^\dagger \ket{\vec{b}}^2 \\
    &\le \max_{\rho_B} \mathop{\mathbb{E}}\limits_{\pairing,U_{\pairing},\vec{b}}  \bra{\vec{b}} U_\pairing \left(\frac{3}{2}\right)^{\nz} \Z^{\otimes \nz} U_\pairing^\dagger \ket{\vec{b}}^2 \\
    &\le \left(\frac{3}{2}\right)^{2\nz}
\end{align}
For a hermitian operator $\op_\circ \propto \canopstring + \canopstring^\dagger$, the final sample complexity can be expressed in terms of the infinity norm of $\op_\circ$,
\begin{align}
    N &\lesssim \mathcal{O}\left(\frac{1}{\varepsilon^2 f} \left(\frac{3}{2}\right)^{\np + 2\nz} \norm{\op_{\circ}}^2_{\infty} \right).
\end{align}